\documentclass[preprint,showpacs,preprintnumbers,amsmath,amssymb,floats]{revtex4}
\usepackage{blindtext}
\usepackage{graphicx}
\usepackage[english]{babel}
\usepackage{amsmath}
\usepackage{amssymb}
\usepackage{color}

\newcommand{\be}{\begin{eqnarray}}
\newcommand{\ee}{\end{eqnarray}}
\usepackage[toc,page]{appendix}

\begin{document}

\title{Anti-symmetric chiral currents at zero magnetic field  \\ in some two-dimensional superconductors} 
\author{Chandra M. Varma}
\affiliation{Physics Department, University of California, Berkeley, CA. 94701 }
\thanks{Visiting Scholar; Emeritus, Distinguished Professor, University of California, Riverside.}
\date{\today}
\begin{abstract}
\noindent
 Non-reciprocal critical currents without applying an external magnetic field have been observed recently in several superconductors, in various forms of Graphene, a Kagome compound and in an under-doped cuprate.
 A necessary requirement for this  is  that the usual supercurrent be accompanied by an anti-symmetric chiral super-current, i.e. with the symmetry of a hall current; equivalently that the superfluid density tensor have an anti-symmetric chiral component. It also requires inversion breaking. The conditions for this phenomena are derived to find that their normal states must break time-reversal and chirality and that the superconducting states must in addition be non-unitary.  Each of the superconductors where spontaneous non-reciprocal critical currents are observed have shown some evidence for such broken symmetries in the normal state. The superconducting state of such materials have topological edge currents, but their projected electro-magnetic part is in general not an integer. The edge states are protected in the superconductor due to a gap. The normal state should show a Kerr effect and, under ideal conditions, an anomalous Hall effect.    
\end{abstract}
\maketitle

\section{Introduction}

Non-reciprocal critical currents in superconductors {\it on applying a magnetic field} and ensuring inversion breaking in a variety of ways, often called the superconducting diode effect, has long been observed  \cite{Ono2020,  Parkin2022, NbSe2_sdiode2022, twistedgraphene_scdiode2022, Yanase2022, Moodera2023} and understood (see for example \cite{Yanase2022, Moodera2023}).  A simple explanation is that it comes from Meissner current due to the applied field, say in a Hall bar geometry, which adds and subtracts to the applied current in opposite directions. The critical current is naturally reduced due to the magnetic field.  The lack of inversion symmetry gives a different value of the net reduction
in changing the direction of the current. At least sometimes, no particular effort appears to be required for inversion symmetry breaking because the couplings to leads on different sides of the sample are never identical \cite{Moodera2023}.  The dimensionless difference in the critical current  applied in one direction and in the other,  $\eta \equiv \frac{1}{2} \frac{I_c^+-I_c^-}{I_c^++I_c^-}$ increases linearly with the applied magnetic field for small fields and so changes sign on reversing the field.

Non-reciprocal critical currents {\it without applying a magnetic field}  have been observed recently in several quasi-two dimensional superconductors, in tri-layer layer graphene \cite{twistedgraphene_scdiode2022}, in the  Kagome compound CsV$_2$Sb$_5$ \cite{LeTian2024}, 
Also J. Wang (Private communication, Dec. 2024), in rhombohedral graphene with four and five layers \cite{LongJu2024_2},  and in the cuprate Bi2212 \cite{JWang2025_CuDiode}. Where studied, the direction of non-reciprocity varies randomly on heating and re-cooling the sample, and if a small external field is applied normal to the sample $\eta$ decreases quadratically in contrast to the usual "diode" effect.  As explained later, the direction of non-reciprocity may be trained by applying and removing a field, both well above the superconducting transition temperature. 
 This spontaneous  effect should be distinguished from the magnetic field induced effect, both because it occurs without applying a field, and because of such differences in characteristics. It should also be distinguished from the non-reciprocity due to Josephson effects in twisted bi-layers in cuprates \cite{PKim2023, Mandar2024, Franz_Jdiode2023}. 

Quite obviously, time reversal (TR)  and inversion  (I) symmetries must be spontaneously broken for the spontaneous diode effect in the compounds discussed here.  These broken symmetries were  predicted in the normal state some time ago in relation to the cuprates due to loop-currents, which lead in the superconducting state to a complex admixture of the
 dominant $d$-wave pairing  to an odd parity state \cite{NgV2004, LNgV2007}.  This alone is not sufficient (because the product of time-reversal and inversion is preserved in the state proposed), but a decoration proposed later to understand properties of the pseudogap state in cuprates is. This state is topological in the normal state \cite{Yiwen2025} and leads in the superconducting state to a chiral topological superconductor. We will investigate also the hexagonal lattice to understand the observation of the same effects in the Kagome superconductor where there has been evidence of time-reversal and chirality in the normal state \cite{teng22n, xing24n, Moll2022, WuLiang2022} due not to any spin-order but to orbital order through loop-currents \cite{tan21prl}. The symmetries in the superconducting state are TR breaking states of d-symmetry states required of superconductivity with Fermi-surfaces around  the $M, M'$ points in the Brillouin zone.  In the rhomobohedral graphene additional symmetries appear to be broken in the normal state \cite{LongJu2024_2}, see also \cite{Zhang-Vishwa}, which suggest a valley-spin polarized state.  A valley polarization usually accompanies an appropriate geometry of loop-currents. There is evidence that small Fermi-surfaces around the ${\bf K}$ points result with full spin-polarization. The other valley and spin are far from the Fermi-surface. Spin-polarization at the Fermi-surface implies odd parity pairing with center of mass momentum $2K$. As discussed later, this uncomfortable situation may be avoided.  A   preprint has very recently appeared on chiral superconducting states for some compounds with emphasis on 
 optical properties of cavities made of chiral superconductors \cite{Ahn_Vishwa} and discussing the general conditions for magneto-chiral conductivity.  Our conclusions  for all the cases we have examined for  spontaneous non-reciprocal critical current are similar. See also \cite{Scammell_2022}. The symmetries in Magneto-chirality, in agreement with these works, and its effects on an observed variation of the ordinary Kerr effect \cite{Kapitulnik1}  was investigated  \cite{AjiHeV, HeLeeV} earlier.

 All the compounds discussed appear to share the feature that the normal state breaks time-reversal, and is chiral.  In two-dimensions chirality does not ensure inversion breaking (or reflection breaking in all planes), and this must be additionally broken. Also, the chirality must be anti-symmetric as in the Hall effect in the normal state.  If one notes that  the Meissner current has the same symmetry as the Hall effect in the normal state, these are the necessary properties of the diode effect of ordinary superconductors in a magnetic field. One need to reproduce these conditions by spontaneous symmetry breaking  to have the spontaneous diode effect.
 
 \begin{figure}
 \begin{center}
 \includegraphics[width= 1.0\columnwidth]{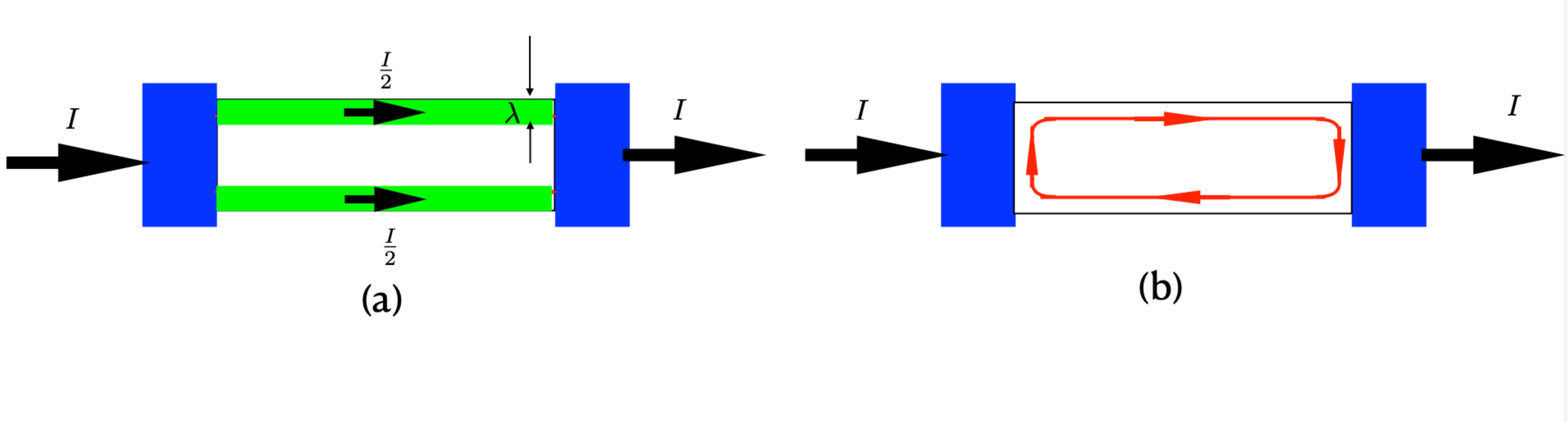}
 \end{center}
\caption{Figure illustrating how the critical currents are measured and how the diode effect comes about. The superconducting sample is connected to very low resistance normal leads through which a current is passed. The usual super-current is illustrated in (a). The anti-symmetric chiral current is due to part of the current at the edge near the leads in the perpendicular direction, opposite at the two edges as shown in (b). With infinite resistance in the normal direction, the current must form a macroscopic closed chiral loop in opposite directions for different directions of imposed current. Inversion breaking leads to different magnitude of the chiral currents for opposite directions of imposed currents. The superconductor turns normal  when the sum of the usual super-current and the chiral super-current at any edge exceeds the critical current.}
 \label{Fig:normal/chiral}
\end{figure}

This is expressed algebraically as follows: The usual static gauge invariant response to a vector potential $A_j$ is a current $J_i$ given by the current-current correlation $\kappa_{ij}({\bf q}, 0) \equiv <J_i J_j> ({\bf q},0)$ in a superconductor with $q \to 0$,,
 \be
 \label{ns}
 J_i &=& \kappa_{ij} A_j, \\ 
 \kappa_{ij} &= &Lim_{q \to 0} \frac{n_s e^2}{m_i}(\delta_{ij} - \frac{q_iq_j}{q^2}).
 \ee
 This form ensures that the response is gauge invariant and transverse. We have noted that the usual response follows the symmetry of the crystal through the effective mass $m_i$ along the principal directions. The first term is the diamagnetic  and the second the paramagnetic effect. This is illustrated in Fig. (\ref{Fig:normal/chiral} - (a)) in the usual experimental method of measuring the critical current by having the superconductor  connected in series to a normal electrodes to which  a current source is connected.  To get a non-reciprocal critical current response, we need beside (\ref{ns}),  a response to an applied current or a vector potential,  in the limit of $q \to 0$ which may be written as 
 \be
 \label{nc}
 J_i^c &=& \kappa^c_{ij} A_j, \\
 \kappa^c_{ij} &=& \frac{n_s^c e^2}{m_i} \epsilon_{ij} (1+ p Z_j).
 \ee
$\epsilon_{ij}$  is the antisymmetric matrix and $Z_j$ is $\pm$ depending on the polarity or  direction of $J_j$. $p$ denotes an amplitude for the polarity, i.e. on the magnitude of inversion breaking or relevant reflection breaking. Such an effect can only arise from the paramagnetic response. $n_s^c$ encodes the broken time-reversal, the associated anti-symmetric chirality, and the broken inversion (or relevant reflection) symmetries. The current $J_i^c$ in the direction orthogonal to the applied current source at the edges where the current enters and leaves the sample is shown in Fig. (\ref{Fig:normal/chiral} - (b)) and must form a closed loop through the two other edges.

While the occurrence of $n_s^c$ spontaneously in the superconducting state is not forbidden in general, it is most natural that (apart from superconductivity) the symmetries encoded in it are spontaneously broken already in the normal state. Such symmetries lead to an anomalous dc Hall effect in the normal state which should have the same symmetries as those that give the Kerr effect. Kerr effect is indeed seen in the cuprates in the pseudogap state \cite{Kapitulnik1, Kapitulnik2} and with some disputed conclusions in the Kagome compound \cite{WuLiang2022}. The dc anomalous Hall effect is not seen in the normal state in either the cuprates or the Kagome compound, while it clearly is in  rhombohedral graphene. The dc anomalous Hall effect in the normal state may be harder to see because it is not protected by a gap as the quantized Hall effect in semiconductors and indeed its inherited properties in a superconductor again due to a gap at the chemical potential. A training procedure whereby such an effect may be seen is discussed in the concluding section.

In a microscopic theory, it is sufficient to show a finite  $n_s^c$ to understand the diode effect. This is done by showing the conditions by which time-reversal and inversion breaking lead to an edge electro-magnetic current. It is shown that such a current is not equal to the quasi-particle or Bogolubov current which is quantized and necessary for the electro-magnetic current, which is always smaller. The square lattice case of the cuprates will be discussed first followed by the hexagonal lattice for Kagome. The case of multilayer rhombohedral graphene will be briefly introduced in a subsequent section and its discussion deferred; it presents some special features of pairing. The relation of the results to experiments and the features of the experiments which need both more experiments as well as more detailed analysis are mentioned in a concluding section. 

  Most of the results here were first reported at the IAS conference at HOKUST (Hong-Kong) (Dec.17, 2024).
 
 \section{Square Lattice - the case of underdoped cuprates}

\begin{figure}
 \begin{center}
 \includegraphics[width= 0.8\columnwidth]{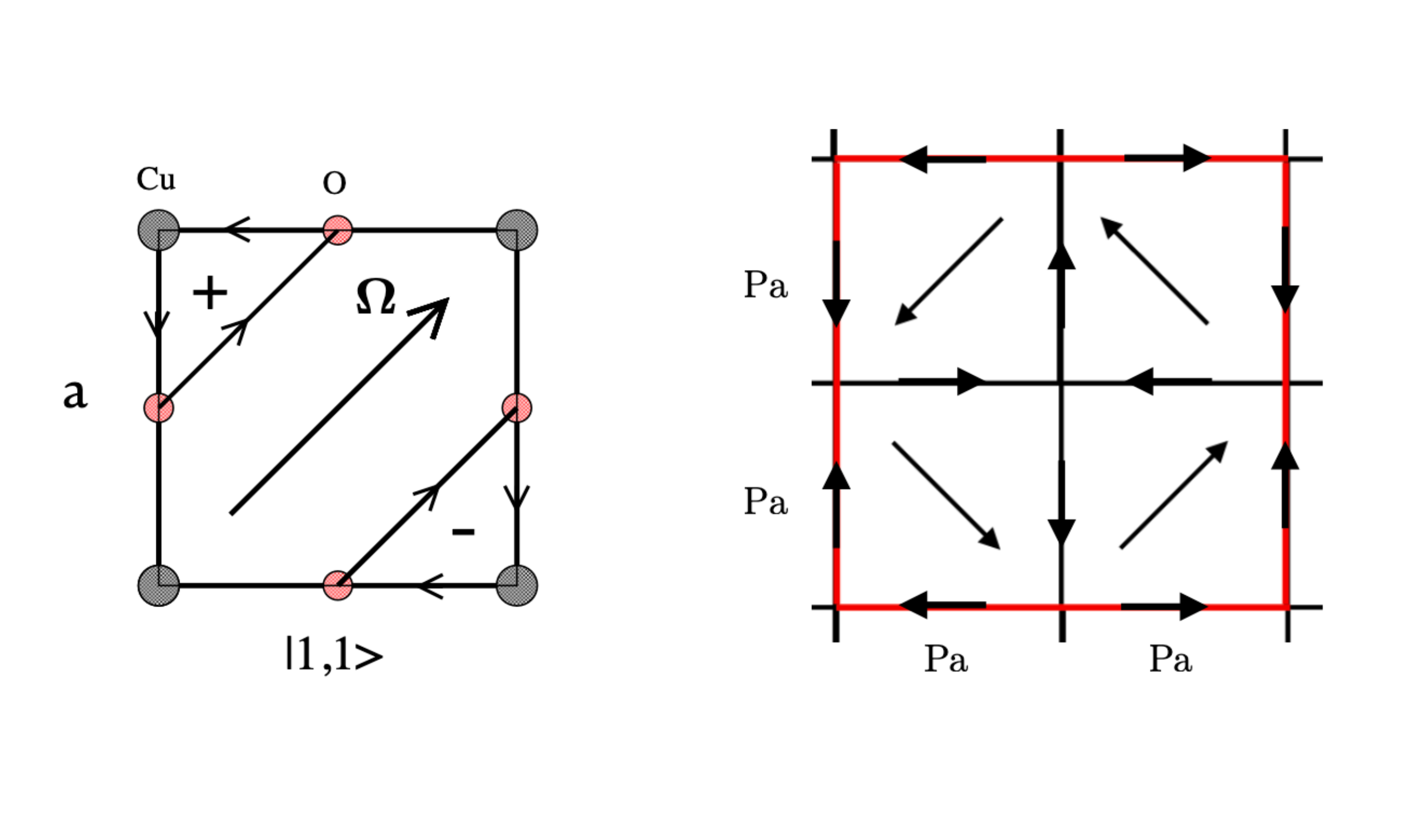}
 \end{center}
\caption{Left: The TR and I breaking order parameter proposed for the pseudogap state of cuprates with loop-currents \cite{simon-cmv}. One of the four possible domains of the order are shown. Since this does not break translation symmetry, it does not give phenomena such as the observed "Fermi-arcs" and the small Fermi-surface observed in magneto-oscillation experiments. Right: To redress this, a decoration of the order with periodic domains arranged periodically in $2P \times 2P$ unit-cells was proposed. To conserve current, the boundaries of the domains have currents which necessarily have vortex arrangements. This order is topological.} 
\label{Symmetries}
\end{figure}

A TR and I breaking phase based on loop-current order was proposed \cite{cmv1997,  simon-cmv} as the symmetry breaking in the underdoped phase of cuprates ending at a quantum-critical point. It was necessary to consider periodic modification of this order \cite{CMV-PLO2019} to understand several properties, for example the Fermi-arcs \cite{ZX-Rev} and the small Fermi-surface observed in magneto-oscillation experiments \cite{SebastianProust2015}. I will first consider the original proposal which is enough to show edge states quite simply and then give the properties of the superconducting state inherited from the modified normal state. 
Both loop-current orders are exhibited in Fig. (\ref{Symmetries}). The modified order gives a finite Berry curvature density in the conduction band which has been calculated recently \cite{Yiwen2025}. This is consistent with the Kerr effect \cite{Kapitulnik1, Kapitulnik2} and Photo-Galvanic effect \cite{Kapitunik_Lim2022} observed in cuprates. The symmetry requirements for  the Kerr effect are the same as those for the dc anomalous Hall effect, which has not been observed. The reason and possible cure for the discrepancy will be discussed in the concluding section. A normal state of such a symmetry is consistent with some recent polarized neutron scattering experiments \cite{Bourges2022, Bourges2023}.

The TR and I breaking in the normal state lead perforce to a modification of the superconducting state from the dominant d-wave state. Consider a two-dimensional TR and inversion breaking superconductor with the order parameter $d(xy)+ i  \epsilon ~ p(x)$\cite{NgV2004}. $\epsilon$ depends on the  normal state TR and I breaking order \cite{cmv1997, simon-cmv}.  (In relation to the cuprates, the $x,y$ axes are chosen at $\pi/4$ with respect to the crystalline axes.)  Given that $i p(x)$ is odd in both time-reversal and in reflection about the y-z plane,  the free-energy contains Chern-Simon terms,
\be
F_{cs} = i \zeta ~ j_y Re(d(xy)^*  p(x)), 
\ee
where $j_y$ is a current in the $y-$direction and $Re(d(xy)^*  p(x))$ is a (superfluid) density; 
 $\zeta$ was derived in \cite{LNgV2007} and depends  on the magnitude  $\epsilon$  and the stiffness coefficients of the superconducting states. I first show here that $j_y$ is a quasi-particle current which may be called the  Bogolubov edge current, which is quantized, and find the magnitude of the corresponding electromagnetic current, which is unquantized. Also shown are the topological aspects of the phenomena by calculating the Chern number for the Bogolubov current. 
 
\begin{figure}
 \begin{center}
 \includegraphics[width= 1.0\columnwidth]{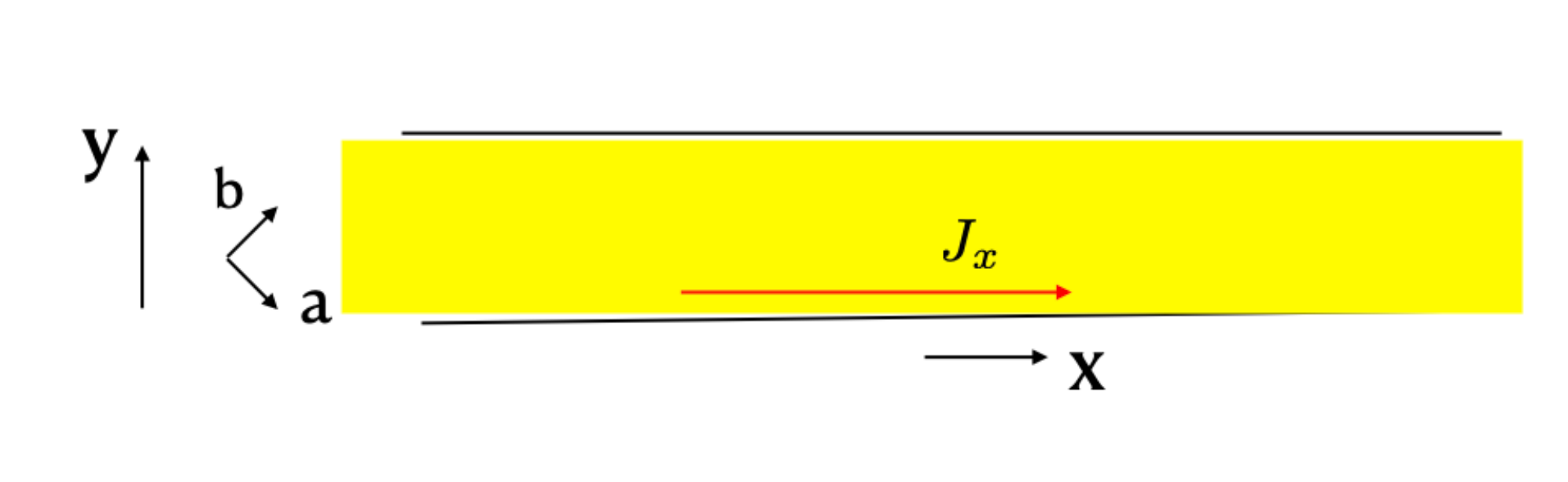}
 \end{center}
\caption{Figure showing the geometry of the calculation of edge-current for the state $d_{xy} + i \epsilon p_x$. An infintely long sample in the x-direction, finite in the y-direction is considered. The state is shown giving a current $J_x$ in the x-direction with the width calculate}
 \label{edge}
\end{figure}

Consider a one effective-band Hamiltonian $H$ in the Gorkov-Nambu space,  
with Pauli matrices $\tau_s$ in particle-hole space, $\sigma_t$ in spin-space. We can write a quadratic Hamiltonian in this space as
  \be
  \label{H-1}
 {\bf H}({\bf k}) =  \sum_{s,t =1,2,3}  \sum_{\bf k} d_{st} ({\bf k}) \tau_s  \sigma_t.
    \ee
 The spin dependence is withheld  to begin with. (One must consider the modifications because  $d_{xy}$ must be a singlet and $p_x$ must be a triplet.) Then $H$ is specified by  $d_s({\bf k}) \tau_s$ alone. The kinetic energy and the real and imaginary part of the pairing are, respectively,
   \be
   d_3({\bf k}) &=& -t (\cos k_xa + \cos k_ya) -\mu, \\
     d_1({\bf k}) &=&  d(xy) = \Delta_d \sin({{k}}_xa)sin({{k}_y}a), ~ d_2({\bf k}) = p(y) =  \Delta_p \sin({{k}}_ya).
     \ee
 $a$ is the lattice constant divided by $\pi$.  (\ref{H-1}) is useful only to derive the current flowing at an edge of a sample infinite in the direction of the current flow as illustrated in Fig. (\ref{edge}). I analyze the solutions near the edge at  $y =0$, following the analysis by Jackiw and Rebbi \cite{Jackiw1976} of the Dirac equation and that used by Volovik \cite{Volovik} and by Read and Green \cite{ReadGreen2000}  for $(k_x + i k_y)$ and $((k_x^2-k_y^2) + i k_x k_y))$ superconductors.  
 
The spectra of (\ref{H-1}) with the specified  $d_r({\bf k})$ has a gap $\mu$ at $k_y = 0$ for any $k_x$.  Near an edge parallel to $\hat{x}$, $k_x$ is a good quantum number and the wave-function is expected to rapidly vary with $y$ near the edge at $y=0$.
Near that edge, one may use $k_y = i\partial /\partial y$, so that
\be
\psi(k_x, k_y) \equiv \Delta_d {{k}}_x{{k}}_y + i \Delta_p {{k}}_y = ( -i \Delta_d {{k}}_x + \Delta_p) \frac{\partial}{\partial y}.
\ee 
A solution is sought of the Bogolubov-di Gennes (BdG) equations at the edge of the superconductor, where there is a discontinuity of the chemical potential, of the form
\be
\label{wavefn}
\psi(k_x, y) = e^{i k_x x}  e^{- \frac{1}{\Delta_p} \int_{0^-}^{0^+} dy' \mu(y')} \psi(k_{Fx}, 0).
\ee
To study the edge state for its topological origin, it is sufficient to take the chemical potential of the supercondcutor  to be $\mu = 0$, and  linearize the dispersion near it. For small enough $k_x$,  $(d_3(k_x) - \mu) \propto (k_x a)^2$ in the dBG equations is ignored compared to the linear term in $k_xa$ in the gap-function.   Near the boundary to vacuum, where there is a discontinuity of the chemical potential $\mu_d$, the BdG equations 
 \be 
 \label{dBG2}
  \left(\begin{array}{cc} -{{E}}(k_x)  & - \mu_d(1 - i \frac{\Delta _d}{\Delta_p} {{k}}_x a) \\ \mu_d(1 + i \frac{\Delta _d}{\Delta_p} {{k}}_xa) & - {{E}}(k_x)\end{array}\right) \left(\begin{array}{cc} u(k_x, y) \\  v(k_x, y)\end{array}\right) = 0,
  \ee
have eigenvalues 
\be
{{E}}(k_x) = \mp \mu_d \sqrt{1+ (\frac{\Delta _d}{\Delta_p} {{k}}_x a)^2}.
\ee
The positive eigenvalue solution for a given $k_x$ is bound to the boundary.
Its eigenvectors are given by
\be
\label{uv}
\frac{u(k_x)}{v(k_x)} = \frac{1 + i \frac{\Delta _d}{\Delta_p} {{k}}_xa}{\sqrt{1+ (\frac{\Delta _d}{\Delta_p} {{k}}_x a)^2} 
}
\ee
The excitations are Majorana if $u/v =\pm 1$ which happens only at $k_x a=0$. For $\Delta_p/(\Delta_d k_xa) << 1, ~u/v =  i$, the eigenstate is that of a chiral fermion because the fermion operator $\frac{1}{2} (c_{\bf k}^+ + i~ c_{\bf -k})$ is a chiral fermion moving in the ${\bf k}$ direction. In general the solutions of the eigenvalue equation (\ref{dBG2}) is a linear combination of $\tau_1$ and $\tau_2$ with $k_xa$ dependent coefficients so that they are neither Majornas nor chiral fermions. 
The electromagnetic current carried by the edge state is $e \frac{k_x}{m} (u_{k_x} u^*_{-k_x}  - v^*_{k_x} v_{-k_x})$. For $\Delta_p/\Delta_d \to 0~, k_xa \ne 0$, this is simply  $e \frac{k_x}{m}$ with corrections of order $\Delta_p/\Delta_d$. This is to be contrasted to that for the edge state of a $(k_x + i k_y)$ superconductor which has 
$u/v = 1$ at all ${\bf k}$, so that the edge state is a Majorana at all ${\bf k}$ \cite{Volovik, ReadGreen2000} and the electro-magnetic current is $0$.  The current carrying state decays perpendicular to the edge with characteristic length $\frac{\Delta_p}{\mu} a$.

\subsection{Topology and the Electromagnetic current}

To study the topological properties, one usually studies the 
Chern number $C_{xy}$. For superconductivity, this is the response to a perturbation  
$$\frac{\partial}{\partial {\bf k}} H({\bf k})\cdot {\bf A},$$
 and may be called the Bogolubov response.
\be
\label{chern1}
C_{xy} \equiv  \frac{1}{2} \sum'_{\bf k} \epsilon_{ijk} \frac{\partial_{x}d_{i}({\bf k})  \partial_{y}d_{j}({\bf k}) d_{k} ({\bf k})}{E^3({\bf k})}.
\ee
$E({\bf k})$ are the differences of the two eigenvalues. The restriction on the sum is to occupied states and here only for $T=0$.  Electro-magnetic fields couple only as 
$$\tau_3 \frac{\partial}{\partial {\bf k}} H({\bf k}) \cdot {\bf A}.$$
This is used by 
 Goryo and Ishikawa (GI) \cite{Goryo_Ishikawa1999}  (see also \cite{Kallin_2016, Lutchyn2008, Brydon2019} for high frequency Hall conductivity), to show that in a gauge-invariant theory, the transverse charge conductivity for a  superconductor with a single normal state band is not given by $C_{xy}$ but by 
  \be
  \label{flux}
 G_{xy}  \equiv  \frac{1}{2}\sum'_{\bf k}  \frac{{d}_{1}({\bf k})\big({d}_{2}({\bf k}) \wedge {d}_{3}({\bf k})\big)- {d}_{2}({\bf k})\big({d}_{1}({\bf k}) \wedge {d}_{3}({\bf k})\big) }{E^3({\bf k}) }, 
  \ee
in which $ A \wedge B \equiv (\partial_x A \partial_y B - \partial_y A \partial_x B)$. On a square lattice, with $d_1(k) = \Delta_d k_xa k_ya ,~ d_2(k) = \Delta_p k_ya,~ d_3(k) = t (k_x^2 +k_y^2)a^2 - \mu$, the integral   over all $k_x, k_y$ in (\ref{flux})  gives zero; but integrating over the quarter space $k_y < k_x < -k_y$ for $0 > k_y > \infty$, one gets $\frac{1}{4}(1-\epsilon)$ while its mirror quarter plane give $- \frac{1}{4} (1-\epsilon)$. $\epsilon$ depends on $\Delta_p/\Delta_d$ in a manner  described below. The value of the integral in the two other quarter plane is $0$. On changing the sign of the imaginary part of the gap-function, one reverses the signs. On changing $\pm i \Delta_p k_x$ to $\pm i \Delta_p k_y$, one interchanges the quarter planes in which one gets finite and zero results. These facts are consistent with the directions of current flow mentioned above for each of the four choices of the complex gap functions \cite{LNgV2007}. Using the lattice dispersions and integrating over corresponding parts of  the Brillouin zones gives the same results.

An order parameter for the normal state with an extended period was suggested \cite{CMV-PLO2019} as a further decoration of the time-reversal and inversion odd order parameter, see the sketch in Fig. (\ref{Symmetries}), to explain properties in the pseudogap phase of the cuprates such as small Fermi-pockets \cite{SebastianProust2015} and "Fermi-arcs" \cite{ARPES-revZX}. Macroscopic experiments, Kerr effect and photo-galvanic effect with circularly polarized light (CPGE) consistent with such a state \cite{Kapitunik_Lim2022}. Polarized neutron scattering experiments have also found results consistent with it \cite{Bourges2022, Bourges2023}. The symmetry of the normal state has an extended period with a point group symmetry preserving the product of time-reversal and four-fold rotation; reflection about any of the planes of the square lattice is not a symmetry. A calculation of the Chern density for this state shows a finite value in the occupied band(s) \cite{Yiwen2025}. The superconducting state inherited from such a normal state reproduces all the edge currents revealed above for the four different states $d_{xy} \pm i p_{x,y}$.

\begin{figure}
 \begin{center}
 \includegraphics[width= 1.0\columnwidth]{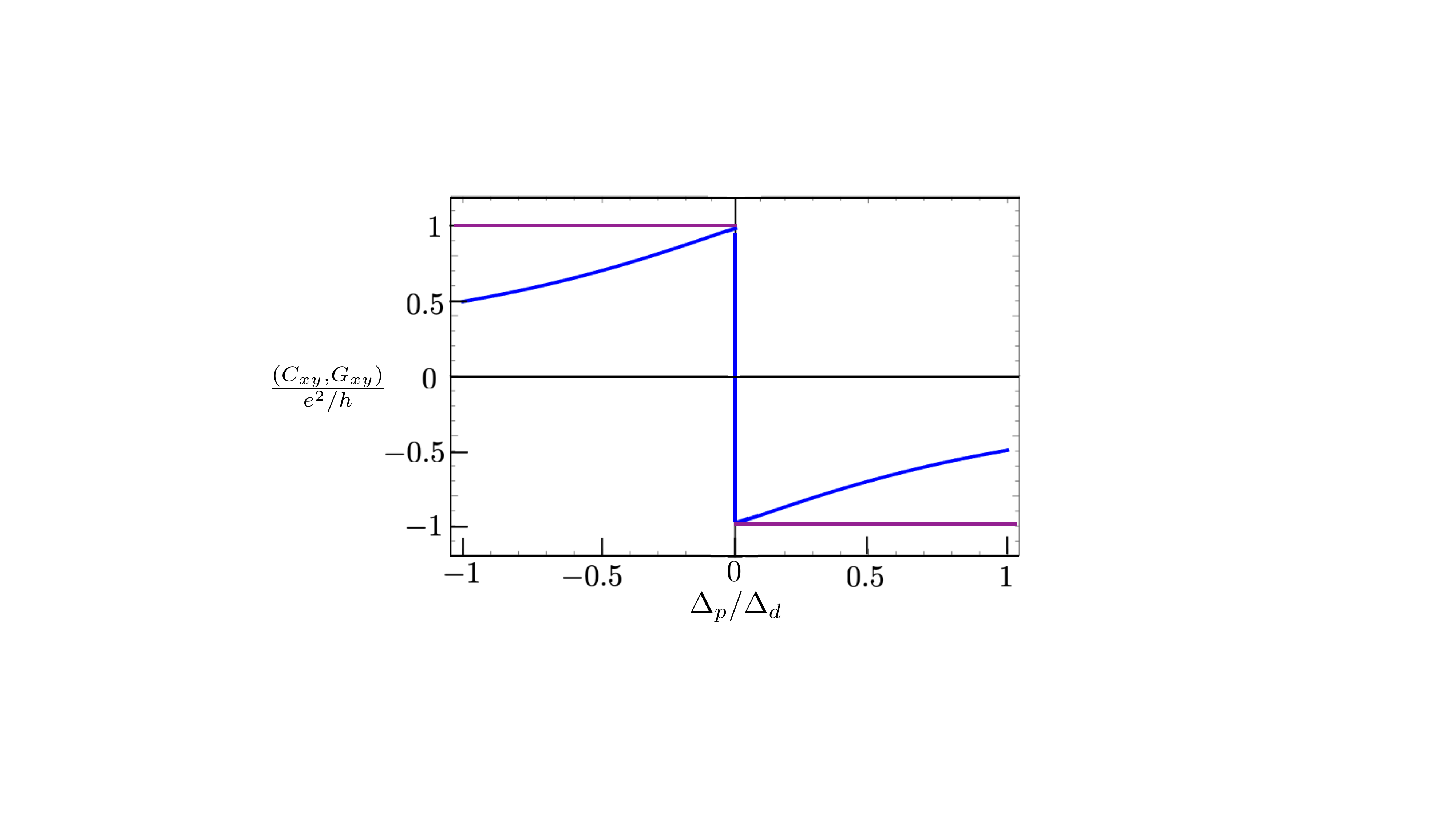}
 \end{center}
\caption{blue: The value calculated of the electromagnetic part of the spontaneous antisymmetric chiral current conductivity $G_{xy}$ normalized to $e^2/h$ as a function of $\Delta_/\Delta_d$. F The calculation used Eq. (\ref{flux}). The Bogolubov edge current conductivity $C_{xy}$ in the same units is $\pm 1$ for all parameters, as shown in purple. }
 \label{sigxy}
\end{figure}

The symmetry of the superconducting state  consistent with such a normal state is
\be
d_1 = \Delta_d k_x k_y; d_2 = \pm \Delta_p {\mathcal R}{\bf {k}},
\ee
 where ${\mathcal R}$ rotates ${\bf k}$ from $k_x$ by $\pm \pi/2$ in going from one quarter or the Brillouin zone to its adjacent quarter. The two values given by ${\mathcal R}{\bf {k}}$ specify the two direction of current flow. Such a state has a gap for all $k_x, k_y$ at $\mu$  unlike the states considered above. On calculating $G_{xy}$ for such a state for the entire k-space, one finds $G_{xy} = (1-\epsilon)$. Note that the anomalous electromagnetic Hall-type conductivity in dimensional units or $n^c_s$ is $\frac{e^2}{h} G_{xy}$. In Fig. (\ref{sigxy}), $G_{xy}$ is plotted as a function of $\Delta_p/\Delta_d$. It is $\pm 1$ for $\Delta_p/\Delta_d \to 0$ from positive and negative values respectively with a value $0$ at $\Delta_p = 0$.  $C_{xy}$ on the other hand is $\pm 1$ for the same values of $\mu$. A calculation of $G_{xy}$ with the periodic band-structure on a lattice and integrating over the Brillouin zone gives the same results near $\Delta_p/\Delta_d = 0$ and slightly different quantitatively elsewhere. Note that the solutions of the BdG equations are not analytic for $\Delta_p \to 0$.

 The physical reason for this remarkable behavior of $G_{xy}$ is to be found above in Eq. (\ref{uv}) and the discussion of edge-current ensuing from it, showing it to be purely electromagnetic, which also explains why it is topological but not quantized (very much like the Meissner current).

Let us now  include the spin-matrices $\sigma_i$. In the expressions  for $C_{xy}, G_{xy}$, one must take the trace over spin.  
 With isotropy in  spin-space, the even parity state is a singlet and the odd parity state a triplet. Consider first that the $S=1, m_S = 0$ odd parity state. It is easy to solve for the spin-dependent $u$ and $v$ given this spin-dependence for $\Delta_d$ and $\Delta_p$. The diagonal component in spin of each of these are then $0$ and the off-diagonal components follow the equation (\ref{uv}). So the conclusions remain unchanged. 
 For the triplet state with $m_S = \pm 1$, $\Delta_p$ is replaced by a matrix with only diagonal components. The solution of $(u/v)$ for the diagonal components and the off-diagonal components separate on ignoring terms of $O(\Delta_p/\Delta_d)^2$.  To that order, since neither  contain both odd and even functions of ${\bf k}$, no edge current of the kind we have investigated above is possible. If the triplet components have a momentum dependence $(k_x \pm i k_y)$, Majorana fermions separately for the up-up and the down-down spin components occur. The general case is a linear combination of the two cases.

\section{Hexagonal Brillouin Zone for Kagome superconductor}

 Properties of time-reversal breaking and chirality have been attributed to the wave-functions of an {\it effective} one band model above, which can only be obtained starting with a multi-orbital model with interactions. The necessity of this  aspect becomes  clearer  
for the hexagonal and Kagome lattices where the multi-orbital description is essential.  Two of the compounds in which the non-reciprocity has been discovered necessarily belong in this category. For the Kagome compound, a chiral normal state order parameter has been detected in a variety of experiments, see for example \cite{jiang22nsr, yin22n, xing24n, Moll2022, LeTacon2024, wang23nrp}. There have been also several calculations suggesting loop-current ordered states, see for example \cite{lin21prb, park21prb, dong23prb, tazai23nc, fu24ax, li24prl}.  
The microscopic details of symmetry breaking in these compounds is not yet known.
But a minimal multi-orbital model is sufficient to show the general requirements for the topological edge excitations and their electromagnetic content.

\begin{figure}
 \begin{center}
 \includegraphics[width= 1.0\columnwidth]{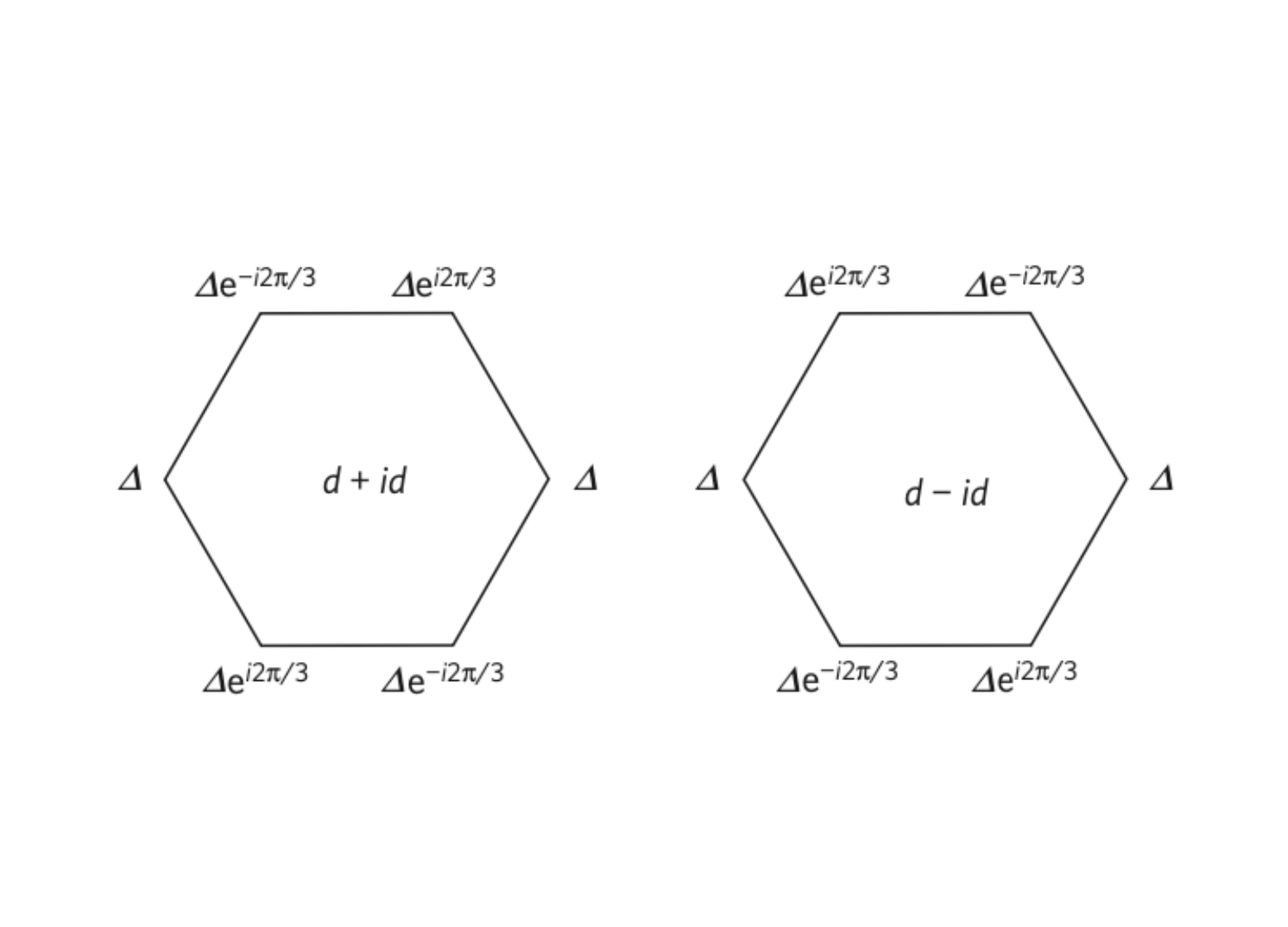}
 \end{center}
\caption{The TR breaking chiral superconducting order parameter due to the two-fold degenerate $d_{x^2-y^2}$ and $d_{xy}$ states on a hexagonal lattice with Fermi-surfaces around the $M$ points which are at the middle of the hexagonal edges of the Brillouin zone(s). The relative phase shift between the orders given by the $\delta$'s in Eq. (\ref{hex-super}) is not shown. Both chiral states are shown. The figures are adapted from those given in A.M. Black-Schaffer, Phys. Rev. B 90, 054521 (2014)}
 \label{hexorder}
\end{figure}

Let us consider  CsV$_2$Sb$_5$, in which measurements of the Fermi-surface in the normal state above superconductivity shows small elliptical pockets centered around the $M$ points. A natural superconducting order parameter (with zero center of mass momentum) if the normal state breaks time-reversal and is chiral is \cite{V-Wang2023}
\be
\label{hex-super}
|\Delta| \Big(m_1 + e^{i (2\pi/3 + \delta_{21})} m_2 +  e^{i (4 \pi/3 + \delta_{32})}m_3\Big).
\ee
where $m_i$ refer to the three pairs of oppositely located pockets with partners of the paired fermions.   With both $\delta =0$, the state is just $(d_{x^2-y^2} + i d_{xy})$ or $(\cos 2 \theta + i \sin 2 \theta)$ due to degeneracy of the $d_{x^2-y^2}$ and the $d_{xy}$ states and a repulsive Josephson coupling between the pairings at different ($M,-M$) points \cite{Nandkishore2012}.  A sketch of the order is shown in Fig. (\ref{hexorder}) with $\delta_{21} = \delta_{32} =0$.
The six fold anisotropy is recognized by evaluating the $\theta$ at Fermi-surfaces near the $M-$ points and keeping zero center of mass momentum. 
$\delta$'s $\ne 0$ are required (and in general $\delta_{21} \ne \delta_{32}$) for the three new principal Bragg spots induced by the normal state transition with varying intensity which can be reversed with an applied magnetic field \cite{jiang22nsr, yin22n, xing24n, Moll2022, LeTacon2024, wang23nrp}. They admix odd-parity state in the superconducting state. We find from calculations that with dominant TR breaking for the d-wave state, they make a negligible change in the Hall conductivity or $n_s^c$

The model Hamiltonian on the basis of the operators which are lattice transform of the operators for the orbitals at the $a_i,~b_i$ with pairing at nearest neighbors $a$ and $b$ sites
\be
H &= \sum_{i, R_{ab}} - \mu  (a_{i,\sigma}^+a_{i,\sigma} + b_{i,\sigma}^+b_{i,\sigma}) + (M - t' sin \phi) (a_{i,\sigma}^+a_{i,\sigma} - b_{i,\sigma}^+b_{i,\sigma}) \nonumber \\ 
&- t (a^+_{i,\sigma}b_{i,\sigma} + H.C.) + 
\Delta_i(a_{i,\sigma}^+b_{i,\sigma'}^+ + H.C.)
\ee
$\mu$ is the chemical potential, the second term is the Haldane topological term with energy difference $M$ between the $a$ and $b$ sites and complex transfer integrals with phase angle $\phi$ between the nearest neighbor atoms on the $a$  sub-lattice and between those on the $b$ sublattice. With a lattice transform to the basis 
 $\big(a_{\bf k \alpha}, b_{\bf  k \beta}, a^+_{\bf - k \gamma}, b^+_{\bf - k \delta}\big)$, where $\alpha,...,\delta = \uparrow$ or $\downarrow$. 
With Pauli matrices $s_i$ in $a-b$ space,  $\tau_s$ in particle-hole and particle-particle  space,  and $\sigma_t$ in spin-space.  We write the quadratic Hamiltonian in this space as
  \be
  \label{H-hex}
 {\bf H}({\bf k}) =  \sum_{i, j, k =1,2,3}  \sum_{\bf k} h_{i j k}({\bf k}) s_i \tau_j \sigma_k.
 \ee
The spin-matrix is again ignored as above; the results on including it are the same as discussed above.  The calculations are then done just with $h_{ij}({\bf k})$.  $\mu =0$ is used and the parameters $M$ and $t$ are adjusted so that there is always a gap at energy $0$ in the superconducting state. For $M < 2 t'$, the normal state has anomalous Hall effect with chern numbers $\pm 1$ for the two bands when one of them is completely filled.

The Hamiltonian in this basis is better visualized as a matrix,
\be
 H_{hex} =  \left(\begin{array}{cccc} h_{33} -\mu & h_{13} & 0 & h_{21}- i h_{22} \\  h_{13}^* & - h_{33}-\mu & - (\overline{h}_{21} - i \overline{h}_{22})^* & 0 \\ 0 &- (\overline{h}_{21} - i \overline{h}_{22})^* & - h_{33} +\mu & h_{13}\\ (h_{21} - i h_{22})^* & 0 & h_{13}^* & h_{33} + \mu \end{array}\right).
  \ee
  The elements in the matrix  are 
  \be
  h_{33}({\bf k}) &=& M - 2 t_p \sin \phi \sum_{nni}\sin ({\bf k. R}_{nni}) \\
  h_{13}({\bf k}) &=& -t \sum_{ni} e^{i {\bf k. R}_{ni}} \\
  h_{21}({\bf k}) &-&  i  h_{22}({\bf k}) = - \sum_{ni} \Delta e^{i (\phi_i + \delta_i  + {\bf k. R}_{ni})} \\
  \overline{h}_{21}({\bf k}) & =& h_{21}(-{\bf k}); ~~\overline{h}_{22}({\bf k}) =h_{22}(-{\bf k})
   \ee
   ${\bf R}_{ni}$ are the three nearest neighbor vectors from $a$ to $b$ sites, and  ${\bf R}_{nni}$ are the three next-nearest neighbors (between $a-a$ and $(b-b)$ sites, on the hexagonal lattice. 
  
  \begin{figure}
 \begin{center}
 \includegraphics[width= 1.2\columnwidth]{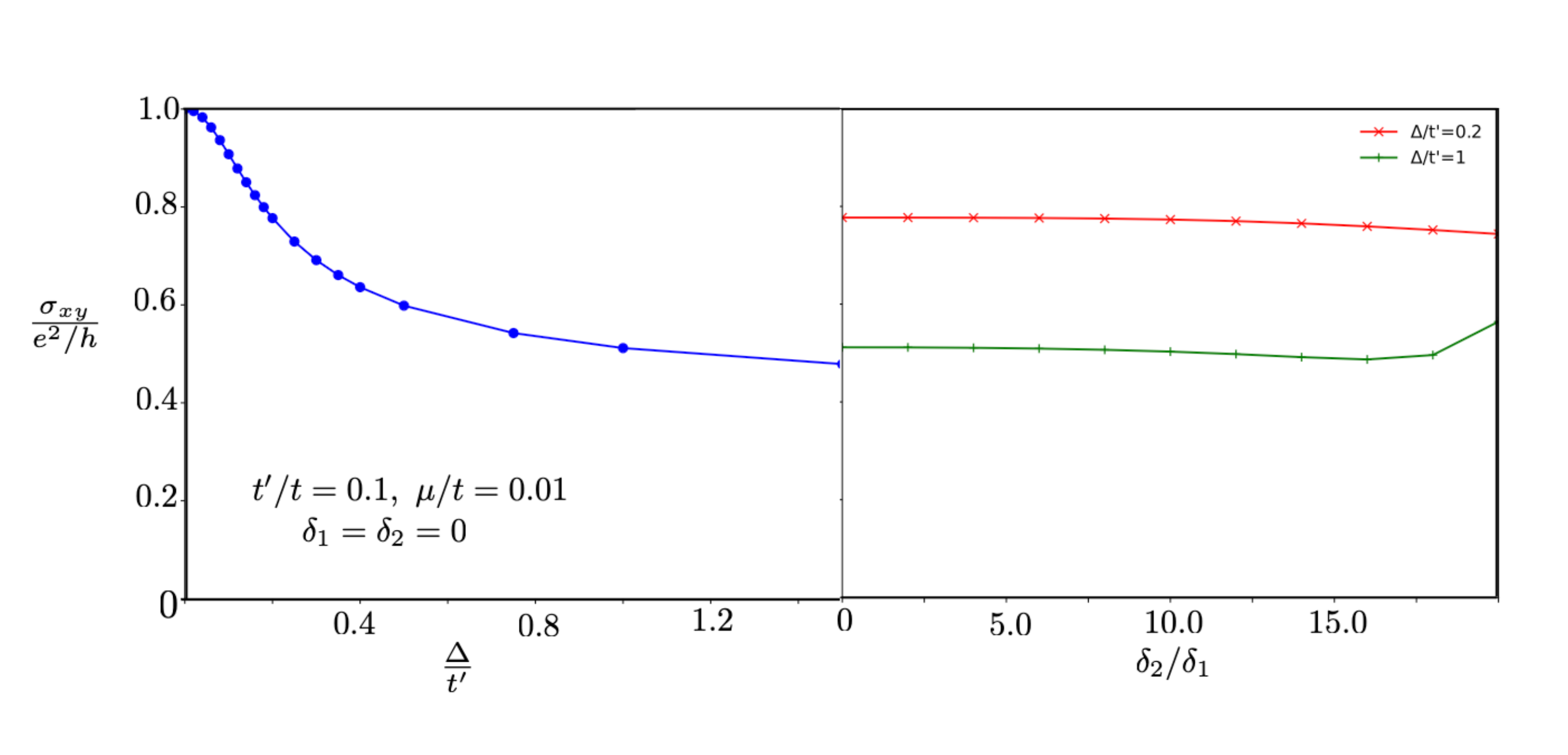}
 \end{center}
\caption{Left: The value calculated of the electromagnetic part of the spontaneous chiral current conductivity $G_{xy}$ normalized to $e^2/h$ as a function of $\Delta_d/t_p$. The calculation used Eq. (\ref{flux}). On the right  the negligible effect of the phase shifts $\delta_{21}$ and $\delta_{32}$ are shown. The Bogoloubov
edge current conductivity $C_{xy}$ in the same units is 1 for all choices, as shown in purple.}
 \label{sigxy-hex}
\end{figure}

Unlike the two band insulator or the one band superconductor, neither the Chern number nor the electromagnetic part of it appear to be expressible through a transformation to the surface of a sphere as  elegantly as Eqs. (\ref{chern1}) and (\ref{flux}).
One can calculate the Chern number for superconductor in the multiband case as in calculating the Hall effect in insulators.  A straight-forward generalization to calculate the electromagnetic current is
gives at $T=0$, 
\be
\label{Gxyhex}
G_{xy} = \frac{1}{2} \sum'_{{\bf k} , m,n} \frac{< m| \partial_{k_x}  \tau_3 H({\bf k}) |n><n| \partial_{k_y}  \tau_3 H({\bf k})|m> -  (interchange ~ \partial_{k_x},~\partial_{k_y})
}{(E_m({\bf k}) - E_n({\bf k}))^2}.
 \ee
$m, n$ are the four bands, and $E_{m,n}({\bf k})$ are their eigenvalues. The prime over the sum indicates that $E_{m}({\bf k}) < \mu, E_{n}({\bf k}) > \mu$. One can also calculate the  quasi-particle or Bogolubov Chern number by coupling all the components of the hamiltonian to the vector potential; i.e. removing the projections to the $\tau_3$ sub-space as in (\ref{Gxyhex}). One gets $\pm 1$, as expected.

The conclusions from numerical calculations of $G_{xy}$ for this case are similar to those arrived at for the single-band problem. The results are shown in Figs. (\ref{sigxy-hex}).
There is no Chern number and therefore no electromagnetic Chern $G_{xy}$ if the topological term in the normal state with $M < 2t' \sin \phi$ is absent. $G_{xy}$ is reduced from integer by the superconducting order parameter $\Delta$ which is of-course responsible for chiral superconductivity. Inversion breaking through the parameters $\delta$ has a minor effect on $G_N$. Inversion breaking through the parameter $M$ makes the pairing amplitudes different at different sites, for example at the $a$ and the $b$ sites of the hexagonal lattice. The superconducting gap in band-space is therefore non-unitary, $\Delta^+ \Delta \ne I$. This has been noticed already \cite{Brydon2019}. Detectable experimental consequences of this need to be investigated.
  
The conditions derived above are all found also in the effective one-band model to get edge currents if we include the fact that the normal state symmetries in the latter are derived from a multi-band model. Only  the minimal effective one band and two band models have been considered above. In reality both the square and the hexagonal Brillouin zones in experiments come from longer periodic translational symmetries.  It is expected that the models investigated have the essential features for the  properties  explored.

\section{Rhombohedral Graphene}

Both four layer and five layer rhombohedral graphene show non-reciprocal critical current in the superconducting state and anomalous Hall effect in the normal state. The nature of the normal state has been discussed both in the experimental papers \cite{LongJu2024_2} and in a recent informative  comment on it \cite{Zhang-Vishwa}. 

The rhombohedral layers are in a large displacement field $D$ and the density of particles $n$ can be varied by gates. A large $D$ leads to valley polarization with a large gap to the next band. There is strong but not definitive evidence of spin-polarization in the occupied valley. The occupied valleys are at $K$ points (or the equivalent $K'$ points) of the hexagonal Brillouin zones as opposed to the $M$ points in the hexagonal Brillouin zones discussed above in connection with the Kagome lattice. The states at the other valley are displaced in energy and may be ignored in the model. Let us take electrons as spin-polarized as suggested by experiments in the normal state. One may take Fermi-surfaces with heavy mass near the $K$ points. (A sketch of the Fermi-surface for a few $D$ and $n$  is given in Ref. \cite{Zhang-Vishwa} near a  K point.)  Superconductivity in such a spin-less, TR and I breaking situation is expected to be  of the $(p_x + i p_y)$ form. The Hamiltonian considered above may be expanded about each of the K points and only with one spin-component. $h_{21}$ and $h_{22}$ in the description above need to be considered. The pairing from any single Fermi-surface pocket is likely to be weird since it carries a large center of mass momentum. The escape from such a situation may be that the difference of momentum between any two K-points is a Bragg vector. Therefore it may be possible to construct states with net zero center of mass momentum by combinations of pairing between particles between the small Fermi-surfaces around different K-points. This is left as a problem to be solved. It is expected that this will not lead to significant difference in the topology leading to a Chern number and its electromagnetic projection.

\section{Magnetic Fields}

 The chiral superfluid density and the resulting edge current are  required by minimization of the Free-energy. We should now consider the magnetic field generated by them (inverse of the Meissner effect problem). The magnetic fields are given by
\be
\nabla \times {\bf B}({\bf r}) = - {\bf B}({\bf r})\frac{R_0}{\lambda^2} + \frac{4 \pi}{c} j_{chiral} \delta({\bf r}).
\ee
$\lambda$ is the London penetration depth and $R_0$ the  order of the circumference of the sample. It is assumed that the coherence length across which the edge current  flows is much smaller than
$\lambda$. The magnetic field then decays over a length scale $\lambda$ with a field at the edge which is just the Biot-Savart value. But what is $\lambda$? In all the samples used, the samples are so thin that the Pearl length is much larger than the transverse dimensions of the sample, but the nominal London penetration depth is much smaller than such dimensions. If one assumes no screening, free vortices are not generated; they can only occur as bound pairs in the superconducting state. It is worth noting that vortices have never been observed in samples whose thickness are on the scale less than 10 nm. This needs to be better understood. It is probably due to such a relation between the dimensions of the sample, the Pearl length and the nominal London penetration depth  that in all the experiments considered in this paper the transition to the normal state by applying a current is abrupt rather than smooth as happens when resistivity due to vortex flow is involved. This  problem is left for further investigations.

  \section{Experiments and concluding Remarks}
  
   In relation to the experiments, given a current source, having a  current flowing around the boundary of the sample together with that in the longitudinal direction achieves the same current configuration as an external magnetic field and asymmetry of right and left electrode of the sample (by spin-orbit coupling  or different roughness or pinning at the two electrodes). These were necessary conditions for the so called diode effect in an applied magnetic field \cite{Ono2020,  Parkin2022, NbSe2_sdiode2022, twistedgraphene_scdiode2022, Yanase2022, Moodera2023}. 
 It is shown in this paper that having spontaneous anti-symmetric chiral currents, i.e. finite $n_s^c$, the same conditions are reproduced without an applied field. In turn, it appears necessary to have  time-reversal breaking in the normal state in a multi-band metal.
 
  Some differences from the effect in an applied field should be noted. On reheating and re-cooling he same sample may produce non-reciprocity in random directions because of the choice of chirality the system may adopt in the normal state. For the same reason,  domains of chirality have effects on the details of the experiments.
  An important distinction is the change of the effect in an applied magnetic field. In the usual diode effect in a magnetic field, the change in the two directions of critical currents is automatically linear in the applied field crossing at zero. We can  use the Ginzburg-Landau theory given in \cite{LNgV2007} to show that the zero-field effect critical current decreases in both directions quadratically. This is as is observed \cite{JWang2025_CuDiode}.
  
   Given a specified current at the edges dictated by minimizing the energy as in the situations above, a magnetic field is generated in the sample to satisfy the Maxwell-London equation (the inverse of the Meissner effect). If the flux  of the field in the sample is small enough compared to a flux quantum, there is no need to solve self-consistently for its effect on the current due to the induced phase difference going around the sample. But 
if the magnitude of the field is such that as a function of temperature, it is several  flux quantums in the sample, the direction of the non-reciprocal current is likely to reverse with temperature as the magnitude of current changes. This may also account for the reversals seen \cite{JWang2025_CuDiode} as a function of temperature as the non-reciprocal current increases. 
It also follows that the effects discussed are discernible easily only in small enough samples. We note that a current of $O(1)$ ~$\mu$-amps on the edge of  a circular sample of 10 micron radius produces at its center a field which if uniform would be about a flux quantum. In rhombohedral graphene, the measured critical currents are about $10$ nano-amps,  so that the induced flux is much smaller than a flux quantum,  in the Kagome compound it is exceeded at the low temperatures while the cuprate compound satisfied this condition only near the superconducting transition where all critical currents are small. These issues are related to those introduced in the previous section and will benefit from further investigations.

Time-reversal breaking and chirality in the normal state have been shown to be essential  for spontaneous edge currents. The striking observation of a non-reciprocal critical current in multi-layer rhombohedral graphene \cite{LongJu2024_2} in samples in which the normal state shows an anomalous Hall effect is 
a clear evidence for the conditions derived in the paper.  In the Kagome compound numerous evidence has been presented in various experiments \cite{jiang22nsr, yin22n, xing24n, Moll2022, LeTacon2024, wang23nrp} for chiral loop-currents in the normal state, but evidence for the anomalous dc Hall effect is missing. But recent experiments at University of Pennsylvania (Liang Wu - private communication Feb. 25, 2025) for high frequency $\sigma_{xy}$, which have the same symmetry as the dc Hall effect appear very promising.) In cuprates extensive evidence \cite{Bourges-rev2023}, with a variety of different techniques, for a time-reversal breaking normal state with inversion breaking and chirality in the pseudogap state is to be found.
Although Kerr effect \cite{Kapitulnik1, Kapitulnik2} at high frequencies and photo-galvanic effects \cite{Kapitunik_Lim2022} in the normal state in cuprates are consistent with the requirements derived above, no evidence for anomalous dc Hall effect has been presented so far. It is worth noting that chiral states at low frequencies 
are not protected as they are for semi-conductors or superconductors with gaps. We suggest looking in the normal state for anomalous Hall effects in small samples and with training in a magnetic field to reduce domains. The field should be applied at high temperatures and removed well above the superconducting transition temperature to measure the normal state Hall effect, as well as the sign of $\eta$ in the superconducting state.

An important prediction for the superconducting state is a macroscopic magnetic field in the sample which may be detected by various techniques, using NV centers or scanning squids, again in small enough samples. The complicated effects of domains should be kept in mind in such experiments. In the normal state of such materials, the magnetic fields should vary periodically in space, as were seen by polarized neutron scattering experiments \cite{Bourges-rev, Bourges-rev2023}. 

There appears much effort at discovering topological superconductors. The easiest experimental method to look for them may be the observation of spontaneous non-reciprocal critical currents.

The pseudogap phase in cuprates below $T^*(x)$ is considered controversial, despite the fact that time-reversal breaking is revealed in a host of microscopic experiments as the only symmetry breaking ever observed starting from $T^*(x)$ and continuing to lower temperatures. The non-reciprocal critical current experiments on the cuprates, since they are macroscopic, should settle this question. It is important therefore that they be independently reproduced in underdoped cuprates.

{\it Acknowledgements}: I owe thanks to Jian Wang in interesting me in the problem and to him and Shichao Qi and Jun Ge for a detailed explanation of the data in cuprates and the Kagome compound, and to Long Ju for the experiments in rhombohedral Graphene. The numerical calculations on the two band model were done by Lijun Zhu to whom I am very grateful. Discussions with Brigitte Leridon, Tai-Kai Ng, and with Aharon Kapitulnik, Rachel Queiroz, Srinivas Raghu, Ziqiang Wang, Nischaal Verma,  and Victor Yakovenko are gratefully acknowledged. Discussions with Beena Kalisky and Amit Keren on field distributions on applying current in samples of various sizes raised some issues which have been mentioned in the concluding section.

%\bibliography{Ref_Oct2024-Edge.bib}
%\end{document}

\end{document}